# Pressure Evolution of a Field Induced Fermi Surface Reconstruction and of the Néel Critical Field in CeIn$_3$

K.M. Purcell,[1,2] D. Graf,[2] M. Kano,[2] J. Bourg,[2] E.C. Palm,[2] T. Murphy,[2] R. McDonald,[3] C.H Mielke,[3] M. M. Altarawneh,[1,2,3] C. Petrovic,[4] Rongwei Hu,[4,5] T. Ebihara,[6] J. Cooley,[7] P. Schlottmann,[1,2] and S. W. Tozer[2]

[1] *Department of Physics, Florida State University, Tallahassee, FL 32306, USA*
[2] *National High Magnetic Field Laboratory, Florida State University, Tallahassee, FL 32310, USA*
[3] *National High Magnetic Field Laboratory, Los Alamos National Laboratory, MS E536, Los Alamos, NM 87545, USA*
[4] *Condensed Matter Physics, Brookhaven National Laboratory, Upton, NY 11973, USA*
[5] *Physics Department, Brown University, Providence, RI 02912, USA*
[6] *Department of Physics, Shizuoka University, Shizuoka 422-8529, Japan*
[7] *Los Alamos National Laboratory, Los Alamos, NM 87545, USA*

We report high-pressure skin depth measurements on the heavy fermion material CeIn$_3$ in magnetic fields up to 64 T using a self-resonant tank circuit based on a tunnel diode oscillator. At ambient pressure, an anomaly in the skin depth is seen at 45 T. The field where this anomaly occurs decreases with applied pressure until approximately 1.0 GPa, where it begins to increase before merging with the antiferromagnetic phase boundary. Possible origins for this transport anomaly are explored in terms of a Fermi surface reconstruction. The critical magnetic field at which the Néel ordered phase is suppressed is also mapped as a function of pressure and extrapolates to the previous ambient pressure measurements at high magnetic fields and high pressure measurements at zero magnetic field.

## INTRODUCTION

CeIn$_3$ is a cubic antiferromagnetic heavy fermion metal[1] that orders with a Néel temperature of 10.1 K and belongs to a family of Ce-based antiferromagnets that exhibit pressure induced superconductivity. A superconducting state arises at a critical pressure $p_c$~2.5 GPa and a $T_c$~170 mK where the Néel temperature is driven to zero.[2] The normal state resistivity at $p_c$ follows a $T^{3/2}$ behavior for T < 8 K that arises from three-dimensional antiferromagnetic fluctuations.[3] It was proposed that the Cooper pairs in the superconducting state are mediated by these magnetic fluctuations.[2,3]

In cerium-based compounds such as CeIn$_3$ the rich physical phenomena are due largely to the competition between the Ruderman-Kittel-Kasuya-Yosida (RKKY) interaction and the Kondo effect. The RKKY interaction induces long-range magnetic order of the f-electrons, while the Kondo effect forms magnetic singlet states between the localized f and conduction states. The respective characteristic energy scales are proportional to $\rho_F J^2$ and $D \exp\left(-\dfrac{1}{\rho_F J}\right)$, where $\rho_F$ and D are the density of states and the bandwidth for the conduction states, respectively, and J is the exchange coupling between f and conduction electron states. The quenching of the f-electron magnetic moments leads to a Kondo resonance and a narrow band characterized by a very large effective mass with heavy fermion behavior.

Doniach[4] proposed that, as a result of the dependence of the RKKY and Kondo energy scales on the exchange coupling J, the RRKY interaction is the prominent energy scale for small values of $\rho_F J$, while for large values of $\rho_F J$ the Kondo effect dominates. When these energy scales are comparable, the critical temperature $T_N$ of the long-range antiferromagnetic order is driven to zero, giving rise to a quantum critical point. A quantum critical transition is a phase transition taking place at zero temperature, i.e. the transition is driven only by quantum fluctuations rather than thermal fluctuations. The magnetic exchange interaction can be varied with the application of pressure, chemical doping or magnetic field. In the case of Ce compounds pressure enhances[5] $\rho_F J$, so that in Doniach's phase diagram the exchange may be replaced by pressure. The antiferromagnetic state of $CeIn_3$ can also be suppressed with the application of a magnetic field.

Ebihara et al.[6] found that at ambient pressure the antiferromagnetic fluctuations are localized in regions referred to as "hot spots" with $\mathbf{k}_{hot}=[\pm 1, \pm 1, \pm 1]k_F/\sqrt{3}$ of the nearly spherical Fermi surface, with Fermi wave vector $k_F$, observed within the antiferromagnetic phase. The position of these "hot spots" corresponds to the location of the protruding narrow necks on the Fermi surface in the nonmagnetic $LaIn_3$. Measuring the angular dependence of the de Haas-van Alphen effect, they concluded that the effective mass associated with the "hot spots" increases with the field as the Néel temperature decreases. Hence, the suppression of $T_N$ with applied magnetic field is accompanied by an increase of the fluctuations in these regions of the Fermi surface, consistent with the presence of a magnetic quantum critical point where the Néel ordered phase is suppressed at ~61 T. Gor'kov et al. proposed that the observed mass enhancement at these hot spots is due to a topological change that occurs in the presence of antiferromagnetic ordering, truncating the necks as they lie close to the boundary of the antiferromagnetic Brillouin zone.[7]

Quantum critical behavior has also been reported[8] in doping studies for $CeIn_{3-x}Sn_x$ for $x_c \sim 0.67$. The Néel temperature decreases with increasing x and at $x_c$ the resistivity has a linear temperature dependence and the Grüneisen parameter diverges with a power law as $T \rightarrow 0$, indicating a quantum critical point. The partial replacement of trivalent In by tetravalent Sn increases the number of conduction electrons and the nearly spherical Fermi surface occupies a larger fraction of the Brillouin zone. Hence, the hot spots on the Fermi surface play a role even in the absence of a magnetic field.

Harrison et al.[9] found further evidence of the suppression of the Néel ordered phase at $H_N = 61$ T at ambient pressure for $CeIn_3$. The Fermi surface cross sections above and below $H_N$ differ, leading to the postulation that the transition is associated with a reconstruction of the Fermi surface. Comparing the Fermi surface of $CeIn_3$ above $H_N$ with that of the nonmagnetic analog, $LaIn_3$, they concluded that the 4f electrons are localized and do not contribute to the Fermi surface volume. This result differs from the high-pressure experiments, in which the 4f electrons are itinerant for pressures outside the boundary of the Néel ordered phase.[10]

Within the Néel ordered phase, Harrison et al.[9] observed the existence of small heavy f-hole pockets, located near the observed "hot spots" of antiferromagnetic fluctuations, that collapse and become depopulated at fields above ~41 T. They proposed that the disappearance of these pockets is associated with a Lifshitz transition[11] and that the formation of superconducting Cooper pairs takes place almost entirely in these f-hole pockets.[12] In contrast to the single transition line expected in the H-P phase diagram for a simple antiferromagnet, our studies show two transition lines for $CeIn_3$, one corresponding to the transition from the Néel state into the paramagnetic phase and a second transition line at lower magnetic fields originating from a reconstruction of the Fermi surface. Below we discuss if this reconstruction of the Fermi surface is the consequence of a Lifshitz transition as proposed in Ref. 12 or a metamagnetic transition.

**EXPERIMENTAL**

We conducted skin depth measurements of CeIn$_3$ in fields up to 64 T and pressures up to 2.3 GPa utilizing a tunnel diode oscillator (TDO).[13] The TDO is a self-resonating tank circuit in which a sample is placed inside the coil or capacitor of an LC circuit and this technique has proven to be very useful for observation of Shubnikov-de Haas (SdH) oscillations and surface resistivity studies of metals in pulsed and DC magnetic fields at ambient pressure.[14,15,16] Two different configurations of the TDO circuit were used in this experiment. In the first configuration the TDO circuit was located on the top of the probe at room temperature using two counter wound coils to transformer couple the low temperature sample coil to the circuit.[17] In the second configuration the TDO circuit was directly connected to the coil and located 18-20 cm above field center in the low temperature region of the probe. The quantity measured is the relative change in the resonant frequency from the zero-field frequency and this is proportional to the change in skin depth ($\Delta F/F \propto -\Delta\delta$), which in turn is directly proportional to the square root of the sample resistivity. This allows for a contactless transport measurement, removing the nonhydrostatic conditions that can arise at the contact points, issues of contact heating and open loop pickup that can arise in conventional high pressure transport measurements. Note that an increase/decrease in frequency relates to a decrease/increase in skin depth, and, therefore in resistivity. Obtaining absolute values from a TDO measurement is challenging, requiring detailed calculations involving the sample and coil geometry, the demagnetization factor and the inductance of the cable used to connect the coil to the TDO circuit. Since we were mainly concerned with constructing the H-P phase diagram, a calibration of the system for absolute values was not undertaken. We, therefore, present the change in the TDO frequency normalized by zero field frequency, $\Delta F(B)/F(B=0$ T).

Single crystals of CeIn$_3$ were grown using the self flux method.[18] Samples were then cleaved or spark cut into discs roughly 300 μm in diameter and 50 μm thick and etched at room temperature with 3:1::H$_2$O:HCl solution to remove any excess indium on the crystal surface. The residual resistivity ratio (RRR) of the samples ranged from 13-17 with a residual resistivity as determined by extrapolation of data above 2 K to T=0 K of ~1 μΩ cm. In order to improve RRR, annealing procedures as prescribed in Reference 6 were attempted for some of these experiments though no increase in RRR was obtained.

These prepared samples were then placed inside a 330 μm diameter three turn coil wound with insulated 56 AWG Cu wire. To avoid excessive sample heating due to the rapid change in magnetic field (20 kT/sec) during a pulse, measurements under pressure were performed in a plastic diamond anvil cell (DAC) using a nonmetallic gasket.[19] Perfluoro polyether oil Fomblin YHVAC 140/13 was used as the pressure-transmitting medium. The pressure was determined *in situ* by comparing the pressure dependent fluorescence of several small ruby chips placed in the gasket hole to an ambient ruby also at the same temperature.[20] Favorable comparison of the FWHM values for the ambient and high pressure ruby confirmed the pressure was hydrostatic. The measurements were performed in $He^3$ refrigerators with base temperatures ranging from 335 mK to 400 mK. Several magnet systems were used: the 60 T generator driven long pulsed (2 sec pulse duration); the 50 T short pulsed (25 msec pulse duration); the 65 T short pulsed (45 msec pulse duration) magnets and the 35 T resistive magnets of the National High Magnetic Field Laboratory (NHMFL). Additional measurements at 1.5 K were performed in the three pulsed magnets housed at the Pulsed Field Facility. The temperatures were measured using a calibrated Cernox thermometer located 2 cm above the DAC and values were confirmed by comparison to the $He^3$ and $He^4$ vapor pressures.

## RESULTS

Figure 1 shows the observed anomaly in the skin depth of $CeIn_3$ at various pressures for T~350 mK. The magnetoresistivity (realized as a decrease in the relative change of frequency) is positive and in agreement with published results[21] up to 15 T. The presented data is the culmination of several experiments using different magnets, probes, pressure cells, samples, coil orientations, circuits and circuitry placement. The observation of the phase transitions across a range of samples and experimental conditions indicates that the behaviors observed are intrinsic to $CeIn_3$

Traces for 0.47, 0.76, and 1.23 GPa were all taken in the 60 T long pulse magnet with the coil axis perpendicular to the magnetic field and are shown in Figure 1A. The 0.92 GPa trace included in Figure 1A is discussed below. The position of the anomaly did not change when the pressure cell was rotated such that the coil axis was parallel to the magnetic field, but was less discernable due to an increased field dependent background, most likely attributable to induced currents in the TDO coil and magnetic screening of the sample. Measurements in the 50 T magnet were performed with the coil axis parallel to the field. Although a single axis rotator was used in the 50 T magnet experiments, the stiffness at low temperature of the 0.018" Gore coaxial cable prevented the DAC from being rotated at [3]He temperatures.

The signal strength of the 50 T magnet measurements was less than that from the 60 T measurements due, in part, to the different coaxial transmission cables used between the low temperature sample space at field center and the rest of the circuit at room temperature. The probe for the 50 T system incorporated a Au-plated stainless steel coaxial cable which had a seven fold increase in loss compared to the Ag-plated BeCu semirigid coaxial cable used in the 60 T and 65 T probes. In addition, a different transformer coupled TDO circuit was used, further decreasing signal. Even with the loss of signal for the 50 T magnet and the larger background due to the orientation of the coil with respect to the magnetic field, the skin depth anomaly is clearly seen. For the purpose of clarity, the measurements taken in the 50 T magnet are displayed in a different panel as Figure 1B. The traces in both Figures 1A and 1B are featureless below the field at which the anomaly occurs aside from low field kinks in the normalized frequency caused by the effect of the magnetic field on the solder connections located at field center and a change in slope in the data taken in the 60 T magnet that is due to a change in the magnetic field ramp rate at 28 T.

We attempted to look for variations in the SdH oscillations above and below the anomaly. An annealed sample was loaded at 0.92 GPa, and to increase sensitivity, the TDO circuit was positioned approximately 20 cm above the coil. While this measurement was also taken in the 50 T magnet, the proximity of the TDO circuit to the sample coil yielded a large signal. This arrangement did not yield any quantum oscillations, however the anomaly was observed and this trace is included in Figure 1A.

Figure 2 shows the location of the anomaly, determined using the peak in the field derivative corresponding to the midpoint of the transition in the raw data, at various pressures and temperatures. Measurements made at 1.5 K show that the position of the anomaly differs little from that measured at 400 mK, although the transition is seen to broaden. Measurements at 4.1 K indicate that the field position of the anomaly may be driven to lower fields with increased temperature, but determination of the location of the anomaly is made difficult as the broadening is more pronounced compared to the lower temperature measurements. The error bars were determined by examining the width of the anomaly. The uncertainty in the measurement of the magnetic field is ±0.5% and in pressure only ±0.025 GPa. These uncertainties are smaller than the data points and are not shown. The inset of Figure 2 displays the relative change in frequency as function of magnetic field for P=1.23 GPa at 400 mK, 1.5 K, and 4.1 K to demonstrate the observed smearing of the transition with increasing temperature. Extrapolating the temperature to zero, the transition corresponds to a decrease in the resistivity seen as a jump in the resonance frequency of the TDO.

Characteristic of the Néel transition we observe a non-hysteretic change of the slope in the normalized resonance frequency of the TDO. Figure 3 shows the relative change in frequency as a function of magnetic field for 1.90 and 2.29 GPa. While both traces show a kink in the relative change in frequency with magnetic field, the overall shapes of the traces are quite different. The 2.29 GPa trace was taken in a BeCu DAC in a 35 T resistive magnet, whereas the 1.90 GPa data was taken in a plastic DAC in the 50 T pulsed magnet. The presence of the surrounding BeCu adds a significant background signal to the measurement. Subsequent measurements taken to fields higher than 50 T at pressures below 1.7 GPa also showed the Néel transition as a second transition at high fields.

In conventional resistivity measurements, the signature of magnetic order is a cusp at the transition point. Since the skin depth is proportional to the square root of the resistivity, a similar feature is expected for the frequency change of the TDO as shown in Figure 3. Ambient pressure skin depth measurements of $CeIn_3$ were performed that confirmed the presence of a noticeable change in slope at 10.1 K as the system was cooled through the Néel transition in zero magnetic field shown in the inset of Figure 3. Upon application of pressure in zero-field, the change in slope when the system is cooled takes place at a temperature that has been identified as the Néel temperature in Ref. 2.

The critical fields at which the antiferromagnetism is suppressed at 400 mK were determined by examining the derivatives of the field traces and used to construct the H-P phase diagram shown in Figure 4 as triangles. The uncertainty in the determination of these points, as well as the uncertainty in measurement of magnetic field and pressure, is smaller than the data point as shown. For completeness, we included the reported high field ambient pressure point at 61 T (Refs. 6 and 9) and the high-pressure boundary point of the Néel ordered phase of Ref. 10. The published phase boundary points are represented by the diamond (Refs. 6 and 9) and the square (Ref. 10). Fitting the data to a power law scaling for the critical field of the form

$$H_N(p) = H_{N,0}\left[1 - (p/p_C)^\alpha\right] \qquad (1)$$

, which is similar to the scaling used in Ref. 6 for the suppression of the Néel temperature with magnetic field, we obtained values for the critical field at ambient pressure $H_{N,0}$ of $60 \pm 4$ T, a critical pressure $p_C$ of 2.55 GPa, and the exponent $\alpha = 4.2 \pm 0.9$. The critical field and critical pressure values are in agreement with the published results.[2,9,10] The Néel critical field at 1.5 K also follows the same trend, albeit at lower fields as expected for higher temperatures. Fitting the Néel phase boundary at 1.5 K using Eq. 1, we obtained values for the critical field at ambient pressure of $56.58 \pm 0.32$ T, a critical pressure of 2.52 GPa, and $\alpha = 4.13 \pm 0.09$.

These results indicate that in the neighborhood of the quantum critical point the critical field varies linearly with $(p_C\text{-}p)$, while for pressure close to ambient pressure the critical field decreases approximately proportional to $p^4$. This differs from the pressure dependence of the Néel temperature, $T_N \sim (p_C\text{-}p)^{0.33}$, but the Néel temperature decreases proportional to $(B_c\text{-}B)$ at ambient pressure close to the magnetic quantum critical point, similar to the $(p_C\text{-}p)$ dependence observed for the critical field close to the pressure induced quantum critical point as described above.

Also shown on the constructed H-P phase diagram in Figure 4 and indicated by circles is the position of the anomaly as determined by using the peak in the field derivative of the relative change in frequency. As the pressure is increased, the field value of the transition monotonically decreases up to 1.0 GPa. Above 1.0 GPa, the transition begins to shift to higher fields and merges into the Néel transition at higher pressures.

Due to temperature constraints during the experiments performed in the 65 T magnet, a larger number of values of the critical Néel field at various pressures were measured at 1.5 K and are displayed together with the field position of the discontinuous transition on Figure 5. The values of the discontinuous transition are within the error of those values measured at low temperatures and follow the same trend with applied pressure as in Figure 4, including the kink in the vicinity of 1.0 GPa. It is worth noting that anomaly is seen to once again merge with the Néel phase boundary, but at a lower pressure due to the decrease of $H_N$ at the increased temperature.

## DISCUSSION

Our measurements clearly show two low temperature transitions as a function of field for pressures below 1.9 GPa and one transition for pressures larger than 1.9 GPa but less than the critical pressure of 2.55 GPa. The high field transition corresponds to the Néel critical field and is seen as a kink in the TDO frequency change with field. The transition at lower fields is temperature dependent. Sharpening of this feature in skin depth as the temperature is lowered indicates it corresponds to a discontinuous drop in resistivity at T=0 K. With increasing temperature, the discontinuity is gradually smeared and with increasing pressure the two transitions merge into one transition above 1.9 GPa.

Based on the changing de Haas-van Alphen (dHvA) frequencies at ambient pressure (heavy f-hole pockets are depopulated as a function of field), it has been postulated[9,12] that the lower field transition is a Lifshitz transition. Lifshitz topological transitions can be categorized as one of two types; the gradual appearance/disappearance of a section of the Fermi surface and the continuous creation/disruption of a neck opening/closing a Fermi surface.

According to the TDO measurements, the change corresponds to a reduction of the resistivity, which for non-interacting fermions is to be interpreted as an appearance (rather than disappearance) of a Fermi surface. On the other hand, it can also be interpreted as a reduction of the scattering rate of the electrons. A traditional Lifshitz transition takes place as a function of pressure due to changes in the overlaps of atomic orbitals, but it can as well be realized as a function of magnetic field as a consequence of the Zeeman splitting. Although the smearing of the transition is consistent with a Lifshitz scenario, the discontinuity is not compatible with a continuous topological change of the Fermi surface. The Fermi surface reconstruction needs to be abrupt, i.e. discontinuous or first order in nature, to be consistent with the TDO measurements. Yamaji et al. have shown that in the presence of interactions in a two dimensional system, a Lifshitz transition can become discontinuous[22] as opposed to the continuous transition one would expect from Lifshitz's consideration of non-interacting systems. As $CeIn_3$ is cubic and isotropic, what effect interactions have on the nature of a Lifshitz transition in this material remains in question. The measurement is performed in a high magnetic field so that Landau quantization of electronic levels is necessary. The high degeneracy of the first Landau level in the new pocket may give the appearance of a discontinuous transition. However, we have not observed quantum oscillations associated with this Landau quantization.

Another possible origin for the anomaly is the effect of a low temperature metamagnetic transition. As the antiparallel spins are gradually polarized by the magnetic field, the order may reach a point before the spins are completely polarized at 61 T in which this change in canting becomes discontinuous. A transition such as this would become sharper as $T \rightarrow 0$ K. One would expect the transition to exhibit hysteresis, but the anomaly shows only a small, temperature independent hysteresis of approximately 0.25 T that could be attributed to experimental artifacts, such as temperature excursions during the magnet field pulse. At first inspection, this possibility also appears unlikely as no discernable jump above the level of experimental noise was reported in previous magnetization studies[6] up to 65 T. Upon close inspection of the magnetization trace of powdered $CeIn_3$ shown in Ref. 6, there is, however, a slight change in slope in the vicinity of 45 T that could correspond to a metamagnetic transition, the effect of which is the discontinuous transition we report. Note that the slop of M(H) decreases around 45 T. Should this be due to the disappearance of hole pockets (Lifshitz transition scenario in Ref. 12) then the slope of M(H) is expected to increase.

The kink in the pressure trend of the transition above 1.0 GPa (Figures 4 and 5) may be attributed to a competition between the Néel and Kondo temperatures. Thessieu et al.[23] concluded from Nuclear Quadrupolar Resonance on the $^{115}In$ sites that above 1.5 GPa in zero magnetic field, the characteristic Kondo temperature $T_K$ is higher than $T_N$. With increasing pressure the hybridization of the f-electron and conduction states becomes larger, and consequently $T_K$ increases faster than the strength of the RKKY interaction. Consequently the f-electrons gradually delocalize and eventually become itinerant through the hybridization. The onset of this electronic crossover from localized to itinerant behavior was confirmed[24] by the observation that above a characteristic temperature T* the nuclear relaxation time was almost independent of temperature. This is interpreted as fluctuations in the Ce-4f moments that are localized above T*. T* steeply increases from ~10 K at 1.9 GPa to ~30 K at 2.65 GPa. It is worth noting that at approximately the same pressure as this crossover from localized to itinerant, the Néel temperature begins to decrease more rapidly with pressure, which the authors of References 23 and 24 attribute to the change in the character of the 4f electrons. This is also consistent with the antiferromagnetic H-P phase diagram presented. The high exponent measured (~4) for the pressure dependence of the critical Néel field is consistent with $H_N$ being relatively pressure independent at low pressures and decreasing much more rapidly in the vicinity of the competing energy scales becoming comparable. A direct confirmation of the character of the 4f electrons requires the observation of quantum oscillations to determine the Fermi surface topology below and above 1.9 GPa.

The observation of quantum oscillatory behavior remained elusive in our experiments despite an attempt to increase RRR using annealing procedures as prescribed in Ref. 6. This may be attributed to a number of circumstances. In general, it is more difficult to observe the SdH oscillations (in the resistivity) in comparison to the dHvA oscillations (in the magnetization).[25] The TDO measurement is not a bulk measurement which makes it even more difficult to observe SdH oscillations. The fundamental frequency at which the circuit oscillated was approximately 25 MHz for the measurements performed. At this frequency and temperature, the skin depth is on the order of 10 μm, thus we are only measuring the outermost 13% of the sample. Damage to this surface that occurred when preparing the samples could further inhibit the observation of quantum oscillatory behavior.

There is also a decrease in the sensitivity of the TDO measurement when the TDO circuit is placed at the top of the probe. The change in frequency that is measured with the circuit in this position includes the inductance added by the 1-2 m of semirigid coaxial transmission line connecting the coil to the circuit. This makes the change in the inductance due to sample properties, such as quantum oscillations, of the small coils used in high pressure cells relatively small compared to the inductance of the transmission line and therefore difficult to resolve. Measurements were also made with the TDO circuit approximately 20 cm above the coil, which increased sensitivity, but still yielded no quantum oscillations. Moving the circuit in this manner yielded a fundamental frequency of ~150 MHz, thus reducing the skin depth to ~4 μm and once again making any damage to the sample surface a concern. Non-hydrostatic conditions could also have affected the observation of quantum oscillations, but comparisons of the FWHM of the ruby situated within the sample space with that of an ambient ruby at the same temperature showed that the pressure medium remained hydrostatic at low temperatures for all pressures. Damage to the coil during pressure loading, such as an electronic short between two of the coil layers, could also cause a decrease in measurement sensitivity. Damage such at this would cause a measurable change in the resonant frequency and resistivity of the sample coil. No such change was seen for any of the measurements presented.

Local sample heating is also a concern. Eddy currents induced in the TDO coil during the magnet pulse as well as the rf current in the coil cause some local heating. The resistance of a typical 3-turn coil 330 microns in diameter made with 56 AWG copper wire is 1.5 Ω. Using a low power BD5 diode, the current through the sample will be approximately 100 μA, dissipating 15 nW of power.

## CONCLUSIONS

In summary, we show the pressure evolution of a discontinuous field-induced transition associated with a reconstruction of the Fermi surface deep within the Néel ordered phase. The trend of this transport anomaly is monotonic until approximately 1.0 GPa, where the pressure derivative changes from negative to positive, presumably due to a change in the character of the 4f electrons from localized to itinerant.[23]

We also present the values of the Néel critical field at 400 mK and 1.5 K and various pressures reveal a previously unseen portion of phase space. The Néel critical field is suppressed by applied pressure by a factor of $1-(p/p_c)^\alpha$ with $\alpha \approx 4$, unlike the manner in which the Néel temperature is suppressed with magnetic field and pressure. The ambient pressure critical Néel field of $60.02 \pm 4.27$ T at 400 mK resulting from the power law scaling differs by less than 2% from the published value of ~61 T and is well within the uncertainty reported.[6,9]

Further high pressure experiments are necessary to determine the character of the 4f electrons at high magnetic fields within the Néel ordered phase and the Fermi surface topology outside of the Néel ordered phase, as well as the effect of applied pressure on the characteristics of these heavy f-hole pockets. A topological transition of the Fermi surface should also be observable in the thermal expansion coefficient and ambient pressure heat capacity measurements. These measurements could be done to determine the origin of this transition.

## ACKNOWLEDGEMENTS

We would like to acknowledge N. Harrison and K. Yang for their helpful discussions. This work was supported by DOE DE-FG52-06NA26193, NSF Cooperative Agreement No. DMR-0084173, and the State of Florida. Advances in the transformer coupled contactless penetration depth technology are credited to Los Alamos National Laboratory LDRD-DR20070013. C.P. and R.H. would like to acknowledge that their work was carried out at Brookhaven National Laboratory that is operated for the U.S. Department of Energy by Brookhaven Science Associates DE-Ac02-98CH10886. P.S. acknowledges the support by the Department of Energy through grant DE-FG02-98ER45707. J.C. would like to acknowledge that his work is supported under the auspices of DOE/NNSA.

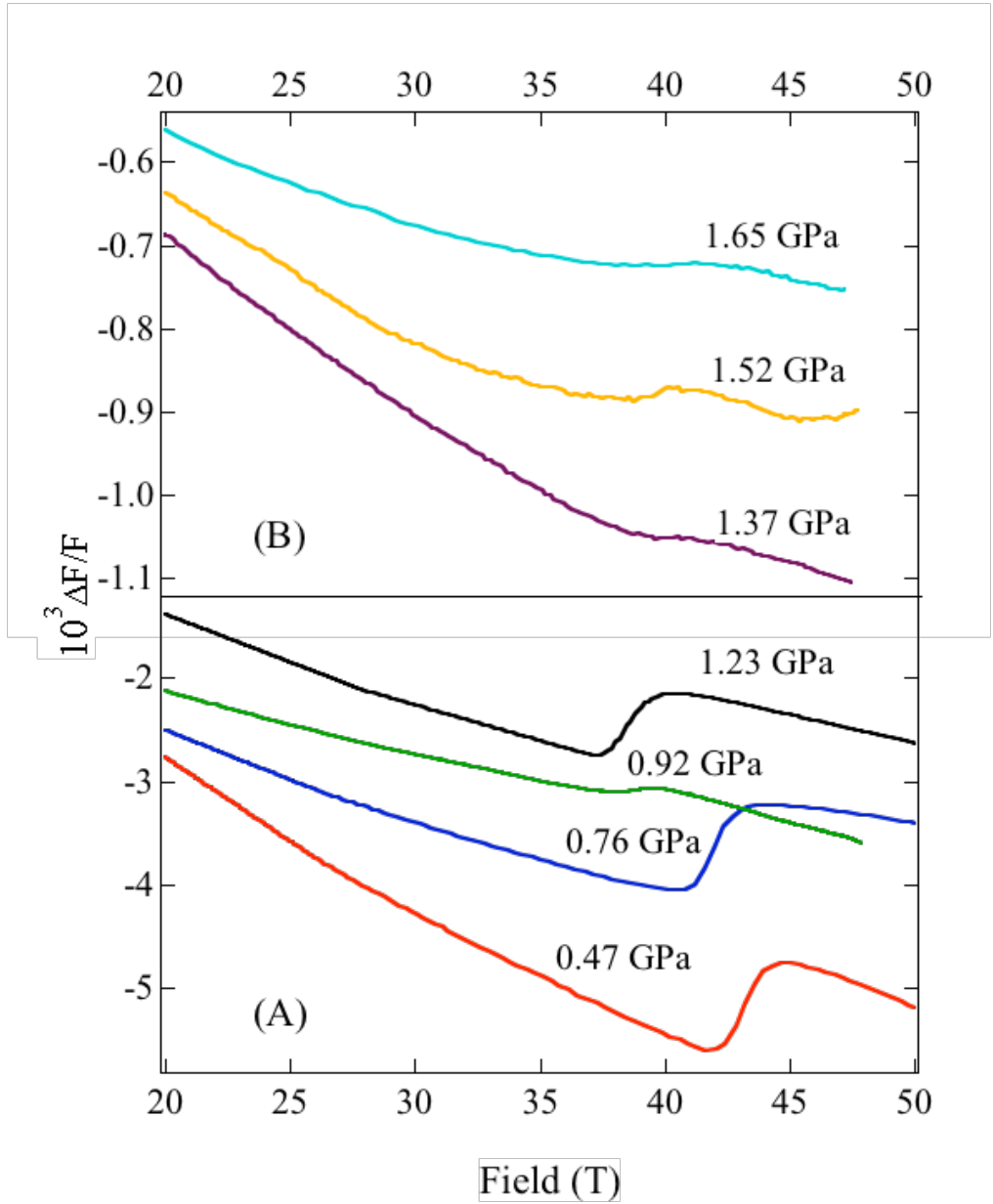

**Figure 1: Relative change in TDO frequency as a function of magnetic field at various pressures. Traces were taken at 335 mK<T<400 mK. Only the traces taken with increasing magnetic field are shown, but the position of the anomaly only slightly changed (0.25 T) on the down sweep. The change in slope seen in (A) at ~28 T is due to a change in the magnetic field ramp rate during the magnet pulse. The measurement at 0.92 GPa was done with an annealed sample in the 50 T magnet. The background of this trace is due to the effect of the magnetic field on the TDO circuit, which was positioned ~20 cm from field center to increase sensitivity in an attempt to observe quantum oscillations. This trace has been shifted by 8.4 $10^{-4}$ for clarity. While no quantum oscillations were observed, the anomaly can clearly be seen. (B) Traces at 1.37, 1.52, and 1.65 GPa had much smaller signal due to experimental conditions discussed in the text.**

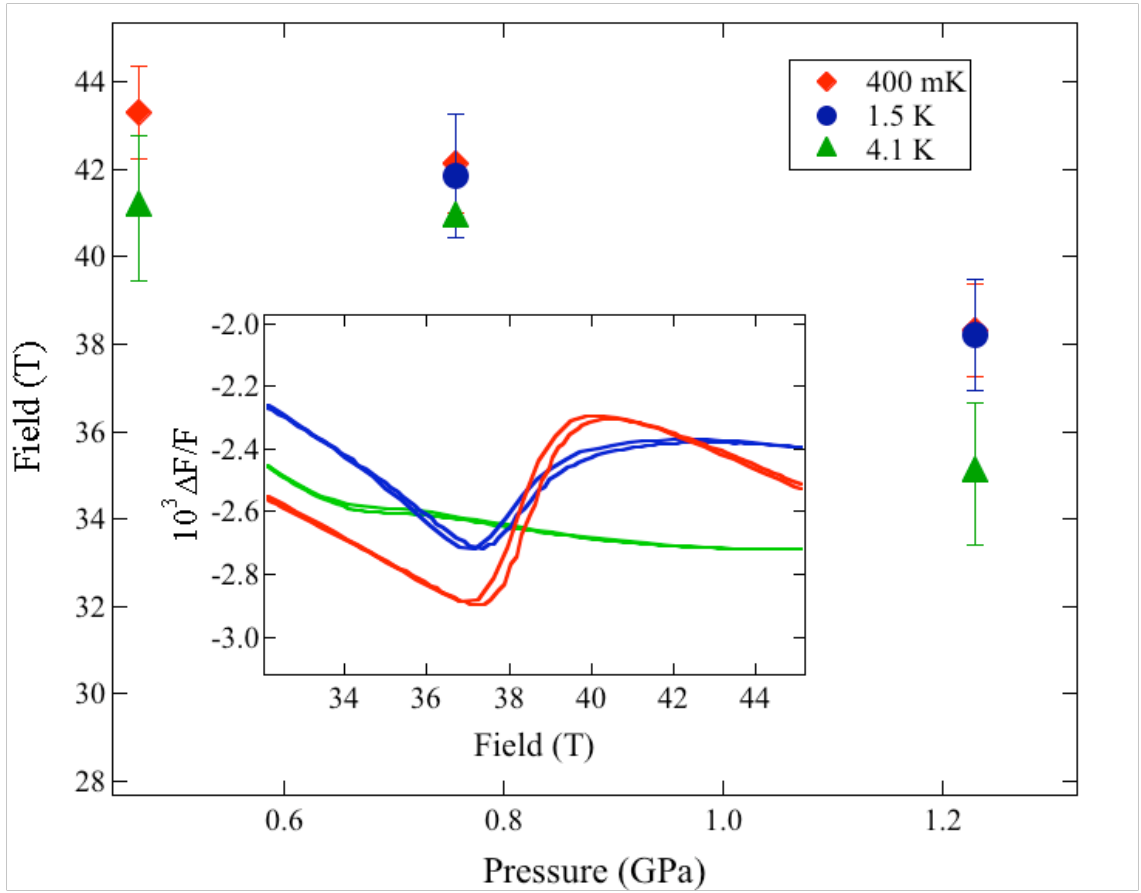

**Figure 2: Position of the anomaly at various pressures and temperatures. The inset is the relative change in TDO frequency as a function of field at 1.23 GPa at 400 mK, 1.5 K, and 4.1 K. The error in the measurement of pressure is ± 0.025 GPa and in magnetic field is ± 0.5 %. Both of these errors are smaller than the data points displayed and errors in the anomaly position determined by examining the width of the transition are displayed.**

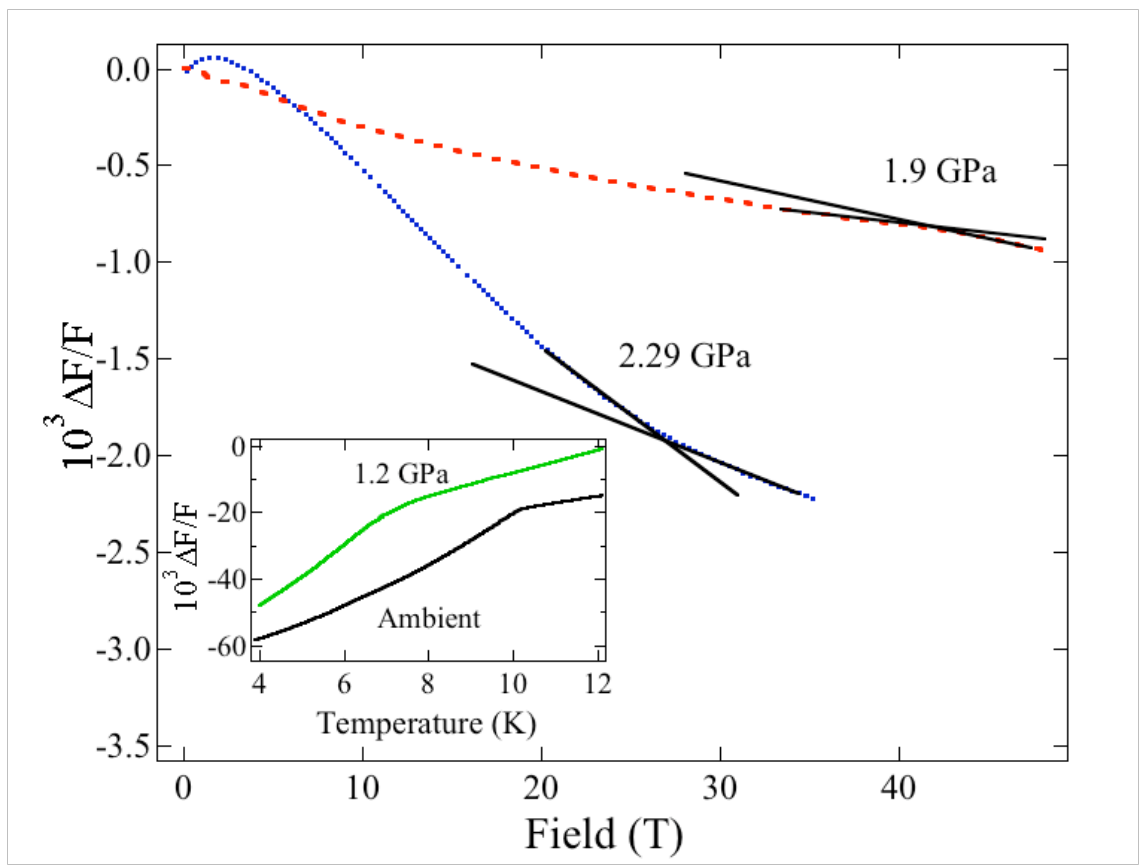

**Figure 3:** Relative change in TDO frequency as a function of magnetic field for 1.9 and 2.29 GPa at T~400 mK showing a change in slope indicating a Néel transition at ~41 T at 1.9 GPa and ~27 T at 2.29 GPa. The background for each of the traces is different as the 1.9 GPa trace was taken using a plastic DAC and the 2.29 GPa in a BeCu DAC and the manner in which the frequencies were mixed differed. The solid lines serve as a guide to the eye, showing the change in slope at the Néel transition. The inset shows the zero field relative change in frequency as CeIn$_3$ is cooled through the Néel temperature at ambient pressure and 1.2 GPa.

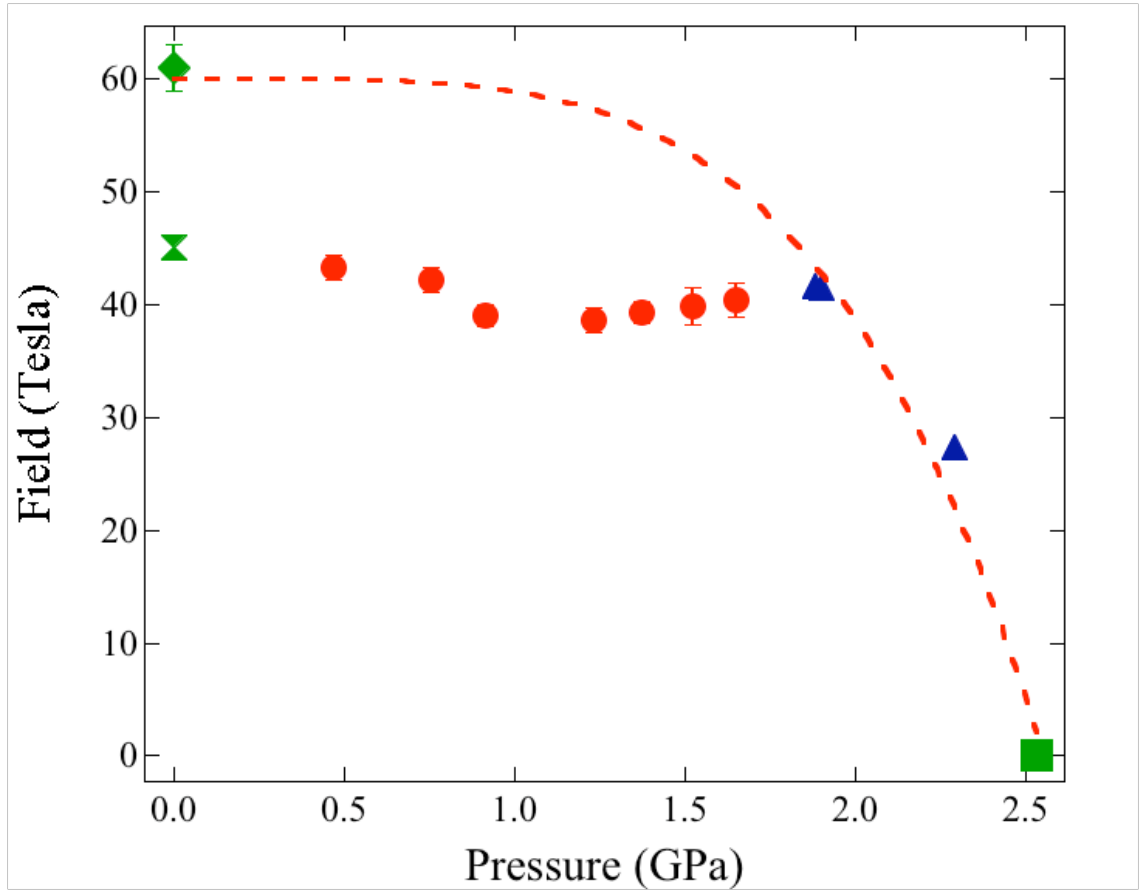

**Figure 4: H-P phase diagram of CeIn₃ at T~400 mK.** The circles indicate the discontinuous transition with errors bars determined by examining the width of the transition. The ambient pressure data point is that observed in Reference 9. The Néel ordered phase is at the border defined by triangles (measured), the diamond (Reference 6) and the square (Reference 2). The error in the measurement of pressure is ± 0.025 GPa and in magnetic field is ± 0.5 %. Both of these errors are smaller than the data points displayed. The dashed line is the fit described by Eq. 1 and yields an ambient pressure critical field of 60 ± 4 T and a zero field critical pressure of 2.55 GPa in agreement the published results.

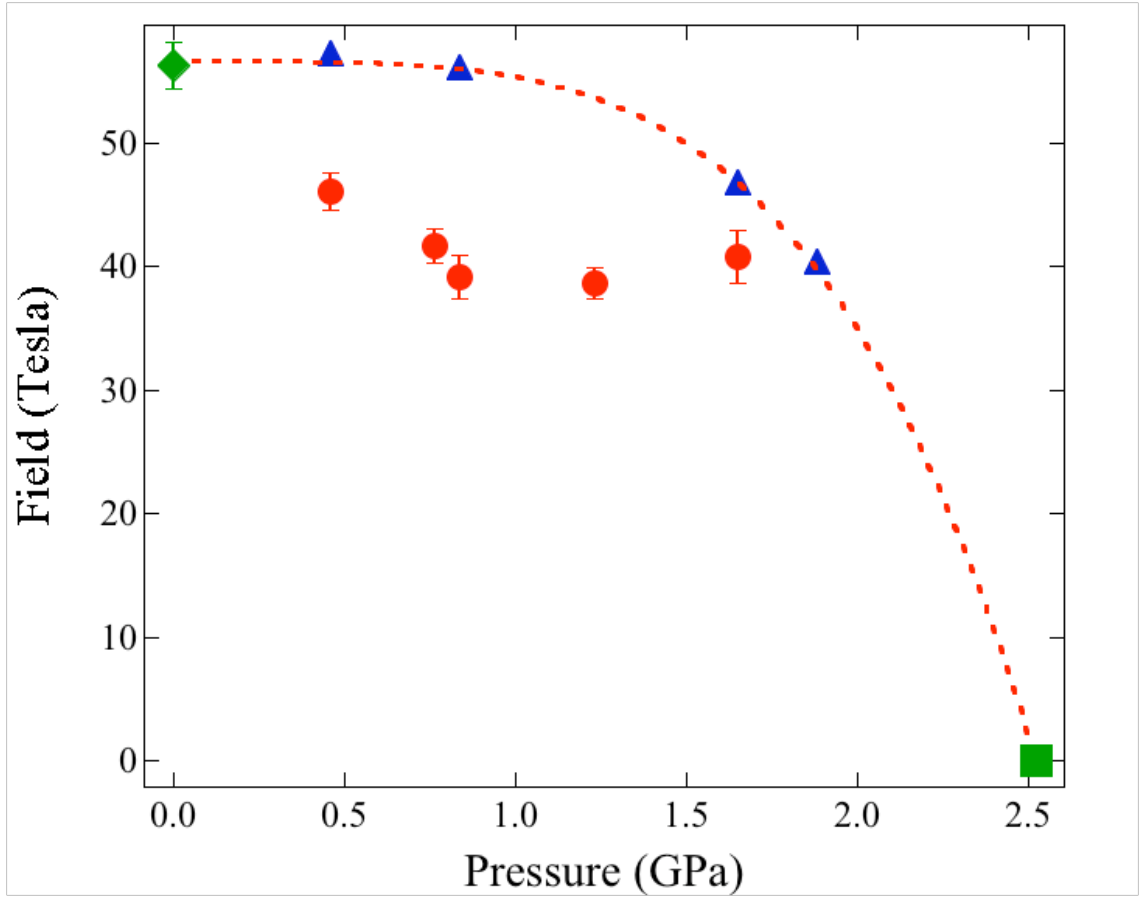

**Figure 5: H-P phase diagram of CeIn$_3$ at T~1.5 K. The circles indicate the discontinuous transition with errors bars determined by examining the width of the transition. The Néel ordered phase is at the border defined by triangles (measured), the diamond (calculated from the fit in Reference 6) and the square (calculated from the fit in Reference 6 of the data from Reference 2). The error in the measurement of pressure is ± 0.025 GPa and in magnetic field is ± 0.5 %. Both of these errors are smaller than the data points displayed. The dashed line is the fit described by Eq. 1 and yields an ambient pressure critical field of 56.6 ± 0.3 T and a zero field critical pressure of 2.52 GPa in agreement the published results.**